# Strongly coupled interface ferroelectricity and interface superconductivity in LAO/KTO

M.D. Dong[1, 2, 3†], X.B. Cheng[1, 2, 3†], M. Zhang[4†], J. Wu[1, 2, 3*]

[1]*Department of Physics, School of Science, Westlake University, Hangzhou 310024, China*

[2]*Research Center for Industries of the Future, Westlake University, Hangzhou310024, China*

[3]*Key Laboratory for Quantum Materials of Zhejiang Province, School of Science, Westlake University, Hangzhou, 310024, China*

[4]*School of Physics, Zhejiang University, Hangzhou 310027, China*

[†]*These authors contributed equally to this work.*

*\*Author to whom correspondence should be addressed: wujie@westlake.edu.cn*

**Abstract:**

**Interfaces can differ from their parent compounds in terms of charge, spin, and orbital orders and are fertile ground for emergent phenomena, strongly correlated physics, and device applications[1-4]. Here, we discover that ferroelectric order resides at the interface of two oxides, $LaAlO_3/KTaO_3$(111) (LAO/KTO), where two seemingly mutually exclusive orders—ferroelectricity and superconductivity—coexist. Moreover, manipulating ferroelectricity can change the interfacial conductivity by more than 1000 times, simultaneously causing superconductivity to diminish and reappear due to the coupling between these phenomena. The ferroelectricity is confirmed by scanning transmission electron microscopy (STEM), second harmonic generation (SHG) microscopy, and piezoelectric force microscopy (PFM). STEM reveals K ions are displaced relative to Ta ions, with**

**the help of oxygen vacancies at the LAO/KTO interface. The resultant electric polarization is locally switchable by applying a voltage between the PFM tip and the LAO film. The ferroelectric hysteresis correlates with hysteresis change in interfacial conductivity and the superconducting transition temperature ($T_c$). The loss and reentrance of superconductivity are accounted for by orders of magnitude change in the Hall mobility induced by the polarization switching. These findings open the door to ferroelectric superconductivity with broken inversion symmetry and non-volatile modulation of superconductivity.**

To overcome the hurdle that the screening of itinerant electrons effectively suppresses the long-range Coulomb forces between electric dipoles and consequently the ferroelectric order, Anderson and Blount proposed in 1964 that ferroelectric metal might exist if itinerant charges are weakly coupled to transverse optical phonons[5]. Since then, experimental evidence of this novel state remained elusive until the ferroelectric metal, $WTe_2$, with a switchable polarization axis, was discovered[6, 7]. So far, the coexistence of electric polarization and metallic conductivity[8] has been observed in several materials, including perovskite oxides[9-11], layered perovskites[12], and two-dimensional materials[6, 7, 13], although the polarization switchability for many of them has yet to be demonstrated.

The lower dimensionality and carrier density reduce electron screening and thus are beneficial for stabilization of ferroelectric order. Besides two-dimensional materials, interfaces of heterostructures, which host a two-dimensional electron gas[14, 15], represent another conducting material system with lower dimensionality and carrier density. It remains an intriguing question whether ferroelectric order might also emerge in these systems, a topic that warrants in-depth theoretical and experimental investigation. Even more attractively, at low temperatures, interface superconductivity has been discovered in heterointerfaces such as $La_{2-x}Sr_xCuO_4/La_2CuO_4$[16-18], $LaAlO_3/SrTiO_3$[19, 20], monolayer

FeSe/SrTiO$_3$[21], EuO/KTO[22, 23], LAO/KTO[22, 24, 25], and other KTO-based heterointerfaces[26-31]. Here, SrTiO$_3$ and KTO are both dielectric materials with large dielectric constants (~ 20,000 and 4,000 at low temperature[32, 33], respectively), indicating that they are on the verge of a paraelectric to ferroelectric phase transition. However, signatures of ferroelectricity in SrTiO$_3$ or KTO-based interfaces remain elusive till now. Electric polarization at the interface, if existing or induced, may be intertwined with superconductivity. Ferroelectric superconductors are exceptionally rare in single-phase materials, with the only known examples being Sr$_{1-x}$Ca$_x$TiO$_{3-\delta}$ bulk[34] and bilayer MoTe$_2$[35]. In contrast, heterointerfaces offer infinite combinations of parent compounds, potentially opening the door to a new family of ferroelectric superconductors.

We chose the LAO/KTO(111) heterostructure for the study of interface ferroelectricity. Both LAO and KTO are insulators, and the electrical conductivity of LAO/KTO(111) comes entirely from its interface. Superconductivity with a transition temperature of approximately 1.7 K resides on the KTO side of the interface[22, 24, 28], and the thickness of the (super)conducting layer is about 4 nm[22, 24], which is less than the coherence length of 18.8 nm[22, 24], illustrating that the superconductivity is two-dimensional. KTO-based interface superconductivity shows an intriguing dependence on substrate orientation; for example, LAO/KTO(111) and LAO/KTO(110) are superconducting, while LAO/KTO(100) is not[22, 23, 25]. This is postulated to be related to pairings mediated by transverse optical phonons[30]. Superconductivity in LAO/KTO(111) can be controlled by back-side gating the KTO(111) substrate, illustrating the dominant role of carrier mobility and interfacial disorder[24]. Thus, the properties at the interface, such as lattice, phonons, carrier density, and mobility, are crucial for understanding interface superconductivity.

A 10 nm LAO film was deposited onto a preconditioned KTO(111) substrate by pulsed laser deposition. During deposition, the substrate was stabilized at 300 ℃ and the oxygen partial pressure was kept at $1\times10^{-5}$ mbar. Water vapor with a partial pressure $1\times10^{-7}$ mbar was added to the growth chamber, a method that has been proven to improve film quality and enhance the superconducting $T_c$. The details of the synthesis are included in the Methods section, and the growth recipe has been continuously optimized through many rounds of synthesis-characterization cycles.

The integrated differential phase contrast (iDPC) STEM images of the LAO/KTO(111) heterointerface clearly show the coexistence of Ta atom displacement and oxygen vacancies right at the interface (Fig. 1). The top LAO film is amorphous, lacking spatial periodicity, in contrast to the regular lattice formed by K, Ta, and O atoms in the KTO substrate (Fig. 1b). The clear boundary between the two materials implies a sharp interface, with cation interdiffusion across the interface being limited to within 1 nm, as determined by energy-dispersive spectroscopy (EDS) mapping of the La, Al, K, and Ta atoms (Extended Data Fig. 1).

K atoms in the vicinity of LAO/KTO interface are displaced relative to the lattice center formed by Ta atoms (Figs. 1c-1e). The K displacement, $\delta$, is along KTO [110] direction and depends sensitively on its distance from the interface. In contrast, the iDPC-STEM image taken from the KTO($11\bar{2}$) plane (Extended Data Fig. 2), which is perpendicular to the KTO($\bar{1}10$) plane in Fig. 1, shows no appreciable Ta displacement. This verifies that the Ta displacement is normal to the KTO($11\bar{2}$) plane and hence it is precisely along KTO [110] direction. The retrieved $\delta$ ($N$) relation (Fig. 1f) shows $\delta$ is huge (~ 0.54 Å) at the interface and decreases rapidly with $N$, where $N$ is the number of TaO$_2$ planes along the KTO [001] direction counting from the interface. The identified positions of oxygen atoms show that a significant number of oxygen atoms in the K-O plane are missing in the KTO lattice close to the LAO/KTO interface, contrasted by the KTO lattice far away from the interface (Figs. 1c-1e). This is corroborated by the change in electron energy loss spectroscopy (EELS) spectra on oxygen K edge taken from area close and far away from the LAO/KTO interface (Extended Data Fig. 3). From the line profile (Fig. 1g), the KTO layers accommodating oxygen vacancies range from $N$ = 1 to $N$ = 6. Thus, the critical thickness at which oxygen atoms are restored to the lattice coincides with the critical thickness at which $\delta$ diminishes (Fig. 1f). This clearly demonstrates that the K displacement is associated with oxygen vacancy.

It should be mentioned that exposure to a high-energy electron beam quickly degrades LAO/KTO samples, especially near the interface, which is susceptible to beam damage, as oxygen vacancies may be created, moved, or removed during STEM imaging. Thus, acquiring atom-resolved images within a very limited time frame is challenging. The STEM images shown in this work are all taken from as-grown samples before severe beam damage occurs.

The electric polarization and the breaking of inversion symmetry come from the K displacement and oxygen vacancies that give rise to nonlinear optical effect detected by the SHG microscopy (Fig. 2a). A linearly polarized laser with 1064 nm wavelength is incident normally on the LAO/KTO(111) sample and it generates a large second harmonic signal with 532 nm wavelength, which is absent in the KTO(111) substrate alone (Fig. 2b). The intensity of SHG signal scales with the laser power $P$ with the expected relation[36] $I_{SHG} \propto P^2$ (Fig. 2c). By rotating the half-wavelength plate, the polarization of incident light is rotated in-plane and the measured SHG intensity $I_x^{2\omega}$ and $I_y^{2\omega}$, with polarization parallel and perpendicular to that of the incident light respectively, show angular dependence on $\phi$ (Figs. 2e and 2f), where $\phi$ is the angle between the polarization of the incident light and the KTO crystallographic $[1\bar{1}0]$ direction. $I_x^{2\omega}(\phi)$ and $I_y^{2\omega}(\phi)$ have 6 peaks in the 360° range with each peak roughly 60° apart that originate from the symmetry of KTO(111) surface. Both $I_x^{2\omega}(\phi)$ and $I_y^{2\omega}(\phi)$ are well fitted simultaneously by expressions based on $m$ polar group, and the retrieved electric polarization is along KTO [110] direction, consistent with STEM results (see the Methods section for details). The symmetry of $I_y^{2\omega}(\phi)$ remains the same and only the amplitude of oscillations changes (Fig. 2g) as the temperature is lowered from 300 K to 200 K. $I_{SHG}$ persists at temperatures as low as 1.6 K, which is below the superconducting $T_c$ (Fig. 2d). This confirms the electric polarization coexists with superconductivity at low temperatures.

The switchability of electric polarization is investigated using the PFM method. On a flat area of LAO/KTO(111), in response to the voltage applied to the PFM tip, we observed a 'butterfly' curve in the switching-spectroscopy PFM (SS-PFM) amplitude, along with a hysteresis loop in the PFM phase, which are both characteristic signatures of ferroelectric order[37-39] (Fig. 3a). It should be emphasized that the application of PFM tip voltage and measurements were carried out in a closed chamber filled with $N_2$ gas. This removes water vapor from environments that breaks down under applied tip voltage and introduces $H^+$ or $OH^-$ ions into the sample[40, 41]. Moreover, any ions induced would have to migrate through the 10 nm-thick LAO layer to reach the interface, making their influence even less likely. In this way, we eliminate the effects due to ion implantation and the measured ferroelectric signal is intrinsic to the LAO/KTO(111) heterostructure.

No ferroelectric domain has been found in the as-grown LAO/KTO(111) under PFM imaging, consistent with the absence of domains in SHG images. However, ferroelectric domains can be created and manipulated by locally switching the electric polarization with the PFM tip[37-39]. By applying a +10 V voltage on the PFM tip, which is sufficient to flip the electric polarization (Fig. 3a), a 4 × 4 μm² square area with polarization opposite to the surrounding as-grown area was created, featuring clear domain boundaries (Figs. 3b-3d). Within this square, a smaller 2 × 2 μm² square was created by applying a -10 V voltage on the PFM tip to flip back the electric polarization. This clearly demonstrates the switchability of ferroelectric domains. Note that the outside as-grown area has the same contrast as the inner smaller square created by the -10 V tip voltage, implying that the as-grown LAO/KTO(111) is in a single domain state, so no domain contrast is present in the pristine state for PFM and SHG imaging.

Combining the results from STEM, SHG, and PFM methods, we conclude that ferroelectricity resides at the interface of LAO/KTO(111), and that this interface ferroelectricity can be switched by an external electric field. Next, we will show the coupling between interface ferroelectricity and interface superconductivity.

We ramp up the PFM tip voltage step-by-step to drive the LAO/KTO through a complete ferroelectric hysteresis loop, while simultaneously measuring the corresponding SHG signal, longitudinal, and Hall resistance to elaborate on the one-to-one correspondence between ferroelectricity and superconductivity. To do this, a designated area between two voltage contacts on a Hall bar is scanned by the PFM tip at a fixed voltage (Fig. 4a), allowing ferroelectricity in this area to be modulated (Fig. 4b). By ramping down the tip voltage $V_{tip}$ to -21 V, then up to +50 V and eventually back to 0 V, the SHG signal $I_x^{2\omega}(\phi)$ and $I_y^{2\omega}(\phi)$ evolves with $V_{tip}$ accordingly (Figs. 4c and 4d). Consistent with Figs. 2d and 2e, $I_x^{2\omega}(\phi)$ and $I_y^{2\omega}(\phi)$ both manifest six peaks with 60° interval. The SHG intensities at the peaks change with $V_{tip}$ but the peak angles remain the same. Since the ferroelectricity has both in-plane and out-of-plane components, and the tip writing only modulates the out-of-plane component, the fittings of $I_x^{2\omega}(\phi)$ and $I_y^{2\omega}(\phi)$ require modeling of sophisticated ferroelectric domain distributions in the intermediate states of hysteresis loop. Nevertheless, putting these details aside, the evolution of $I_x^{2\omega}(\phi)$ and $I_y^{2\omega}(\phi)$ with $V_{tip}$, corroborated with PFM imaging of ferroelectric domains (Fig. 4b), undoubtedly further rules out the possibility of ion implantation by the tip writing and verifies its effectiveness in modulating ferroelectric domains.

The longitudinal resistance $R$ concomitantly shows clear hysteresis behavior (Fig. 4e) that is directly linked to ferroelectric hysteresis loop. It should be mentioned that after the writing process, the effect of tip writing decays over time, quite significantly for $V_{tip}$ < -10 V. Using $R$ as an indicator, the effect decays rapidly in the first 1 minute and then decays gradually on a timescale of hours until it stabilizes (inset of Fig. 4e). Due to the time cost in transferring sample and setting up the optics, the SHG measurements were performed 30 minutes after the tip writing when the quick decaying period was over. $R$ in Fig. 4e was measured *in situ* about 3 seconds after the writing to reflect the maximized effect. The sharp increase or decrease of $R$ at $V_{tip}$ ~ -17 and 6 V respectively, corresponds to critical voltages to flip ferroelectricity. Remarkably, the change in $R$ reaches more than $10^5$ times for two saturated states of hysteresis loop at 300 K, implying a drastic effect of ferroelectric polarization orientation on interfacial conductivity. More importantly, compared to the pristine LAO/KTO sample with $T_c$ = 1.8 K (Here $T_c$ is denoted as the midpoint of superconductivity transition), the superconductivity is weakened at $V_{tip}$ = -17 V and diminishes completely at $V_{tip}$ = -21 V. As $V_{tip}$ reverses its sign and flips the ferroelectric polarization back to the pristine state at $V_{tip}$ = +50 V, the superconductivity recovers (Figs. 4f and 4g). The writing cycles and consequent the reentrance of superconductivity are repeatable, which clearly demonstrates the non-volatile modulation of superconductivity by ferroelectricity.

A comparison of Hall effects in the low/high resistance states reveals the underlying cause for the giant resistance change and modulation of superconductivity. Since ramping the magnetic field takes hours, the Hall effect measurements were performed 10 hours after the writing when a stable state had been reached. During this time, the effect decays and the change for the high/low resistance state is reduced to roughly 355 times. The carrier mobility, $\mu_{Hall}$, retrieved from the measured longitudinal resistance and Hall coefficient, is 261 times smaller in the non-superconducting high resistance state than in the superconducting low resistance state (Fig. 4h). In contrast, the carrier density, $n_{2D}$, changes by a factor of 1.36 in the meantime (Fig. 4i). Thus, the reduction in Hall mobility is undoubtedly the prime reason for the loss of superconductivity.

This is akin to the electric field control of interface superconductivity in LAO/KTO(111) by the backside gating method[24], where the modulation of superconductivity is also ascribed to roughly 10 times change in $\mu_{Hall}$. Here, $\mu_{Hall}$ is affected by the interfacial disorder density and the thickness of conducting layers, both of which are controlled by the gating voltage. Thus, it can be inferred that the disorder density increases substantially during the flipping process of ferroelectricity by the PFM tip. On the other hand, $R(T)$ for the high resistance state, however, has a positive slope from room temperature to low temperatures, characteristics of a metallic state. At first glance, this may seem to be contradicted by the Anderson localization mechanism, where charge localization occurs once the charge mobility falls below the mobility edge. The solution to this dilemma is that the ferroelectric order is within 1.9 nm of KTO/LAO interface (Fig. 1), which is thinner than the conducting LAO layer[22, 24]. The conductivity of the high resistance state stems from the residual conducting layer that is not directly affected by the ferroelectric flipping. Besides the inhomogeneity in the out-of-plane direction, the in-plane inhomogeneity may allow percolative paths for current flow while suppressing the conductivity in most of the area. As a result, the resistance is high but exhibits metallic temperature dependence.

The tremendous amount of disorder in the high resistance state is caused by the flipping of ferroelectric order. Despite the intentionally slow scanning speed and repeated writing processes on the same area, the flipping of ferroelectric order is inhomogeneous at the microscopic scale, leaving behind small clusters unchanged or less affected while scanning through the whole area. This is inevitable since the motion of the PFM tip is not really continuous and there are always gaps between two successive line scans. The PFM signal is a measure of average ferroelectric response that lacks the spatial resolution to resolve these small clusters. One piece of supporting evidence for the above scenario is the significant resistance decay with time after the writing process (inset of Fig. 4e). It implies that the post-writing LAO/KTO interface is metastable and gradually relaxes to a more homogeneous state with less disorder and lower resistance. During the relaxation process, some ferroelectric clusters are flipped back to their pristine state, presumably driven by local tension at the interface due to the inhomogeneous distribution of atomic displacements. Since the atomic displacement is closely associated with oxygen vacancies, the flipping of ferroelectric order is accompanied by the motion of oxygen vacancies. Thus, the timescale for stabilization reaches hours. After completion of one hysteresis loop and a return to the low resistance state, superconductivity recovers but with a lower $T_c$, evidencing that the effect of writing is reversible, and disorder is removed once ferroelectricity flips back to the pristine state. The lowering of $T_c$ is due to a small portion of ferroelectric clusters remaining flipped at the end of the loop. Thus, the ferroelectricity-related disorder scenario provides a natural explanation for our major experimental results.

The interface ferroelectricity and its strong coupling with interface superconductivity are universal for synthesized LAO/KTO(111) samples. More examples from other samples are shown in Extended Data Figs. 4 and 5. The coercivity and shape of the hysteresis loop vary slightly from sample to sample, but the conclusions remain the same.

The discovery of interface ferroelectricity significantly extends the families of materials hosting ferroelectricity and its easy tunability due to lower dimensionality is beneficial for device applications. The coupling between interface ferroelectricity and interface superconductivity engenders non-volatile modulation of superconductivity and fuels research on superconductivity without inversion symmetry.

## Methods

### Film synthesis and device fabrication

A 248 nm KrF excimer laser was used with a 0.5 - 0.7 J cm$^{-2}$ laser fluence and 10 Hz laser repetition rate to deposit 10 nm LAO films onto KTO(111) substrates by PLD. The sample temperature was maintained at 300 °C during growth, in a mixed atmosphere of $1\times10^{-5}$ mbar $O_2$ and $1\times10^{-7}$ mbar $H_2O$ vapor. For the Hall bar used for electrical transport measurements in Fig. 4, we employed standard UV-lithography and lift-off techniques to pre-pattern the KTO substrate and then deposit LAO films to form the desired device.

### Scanning transmission electron microscopy (STEM)

A dual-beam focused ion beam instrument (Helios 5 UX, Thermo Fisher Scientific) was used to prepare a sample of LAO/KTO(111) approximately 30 nm thick. The sample was cut by 30 kV Ga ions and cleaned by 5 kV and 2 kV Ga ions, followed by Ar$^+$ ion milling for 10 minutes at 500 V in a STEM specimen preparation system (Nanomill 1040, Fischione) prior to STEM imaging. The atom-resolved STEM images were collected by a double spherical aberration-corrected STEM (Spectra Ultra, Thermo

Fisher Scientific) operated at 300 kV. The chemical composition was resolved with an energy dispersive X-ray analyzer (Ultra X) in STEM spectrum imaging mode. The STEM–EELS experiments were performed using a Gatan GIF Continuum K3 HR/1069 HR system with an accelerating voltage of 300 kV.

**PFM and SS-PFM**

PFM measurements were carried out using a commercial Environmental Atomic Force Microscopy system (Cypher ES, Oxford Instruments) at room temperature in an $N_2$ atmosphere. The PFM and the switching-spectroscopy piezoelectric hysteresis loops were measured in PFM dual-AC resonance tracking mode. A triangular voltage waveform with a frequency of 0.025 Hz was applied to collect both the amplitude and phase signals of modulated tip vibration as a function of the bias voltage, with a maximum of 20 V. The writing of ferroelectric domains was accomplished by scanning the PFM tip over a designated area with a fixed voltage in the nitrogen gas environment to eliminate ion implantation.

**SHG**

A confocal microscope (WITec, Alpha300RAS) equipped with 1064 nm laser excitation (NPI Rainbow1064 OEM) was employed for non-linear optics measurements (Fig. 2a). The incident light passes through a linear polarizer and a half-wavelength ($\lambda/2$) plate before striking the sample normally. The generated second harmonic signal with 532 nm wavelength passes through the $\lambda/2$ plate and then the second linear polarizer (analyzer). The $\lambda/2$ plate is rotated in 10° step for studies of angular dependence of SHG. For low temperature SHG measurements, the sample is mounted onto a sample holder cooled by an open-cycle liquid nitrogen cryostat capable of reaching temperatures as low as 200 K.

Suppose the polarization of the incident light makes the angle $\phi$ with respect to the $KTaO_3$ $[1\bar{1}0]$ direction, i.e., the $x$ axis, then the electric field is given by

$$\begin{bmatrix} E_x \\ E_y \\ E_z \end{bmatrix} = \begin{bmatrix} E_0\cos\phi \\ E_0\sin\phi \\ 0 \end{bmatrix}$$

The SHG d matrix for the $m$ point group with electric polarization in $(1\bar{1}0)$ plane is

$$d_m = \begin{pmatrix} 0 & 0 & 0 & 0 & d_{15} & d_{16} \\ d_{21} & d_{22} & d_{23} & d_{24} & 0 & 0 \\ d_{31} & d_{32} & d_{33} & d_{34} & 0 & 0 \end{pmatrix}$$

Hence, the SHG signal is given by

$$P^{2\omega} = d_m \begin{pmatrix} E_0^2\cos^2\phi \\ E_0^2\sin^2\phi \\ 0 \\ 0 \\ 0 \\ 2E_0^2\cos\phi\sin\phi \end{pmatrix} = E_0^2 \begin{pmatrix} -2d_{16}\sin\phi\cos\phi \\ d_{21}\cos^2\phi + d_{22}\sin^2\phi \\ d_{31}\cos^2\phi + d_{32}\sin^2\phi \end{pmatrix}$$

Since the SHG light goes through a λ/2 wave plate before it reaches the analyzer, the SHG light intensities corresponding to $x$- and $y$-polarization of the analyzer become

$$I_x^{2\omega} \propto E_0^4 (d_{21}\sin\phi\cos^2\phi + d_{22}\sin^3\phi - 2d_{16}\sin\phi\cos^2\phi)^2$$

$$I_y^{2\omega} \propto E_0^4 (d_{21}\cos^3\phi + d_{22}\sin^2\phi\cos\phi + 2d_{16}\sin^2\phi\cos\phi)^2$$

The above expression fits nicely the SHG data shown in Fig. 2.

**Electrical transport measurements**

Longitudinal resistance and Hall effect measurements were conducted using the Hall bar pattern shown in Fig. 4a. The width of the Hall bar is 30 μm, and the distance between two contacts for longitudinal resistance measurement is 100 μm. The excitation current is set at 100 nA. All electric measurements were performed using a physical properties measurement system (PPMS DynaCool 12T, Quantum Design) with variable sample temperatures ranging from 1.8 to 300 K. For temperatures below 1.8 K, a dilution refrigerator attachment (Quantum Design) is used to reach temperatures as low as 100 mK.

**Data Availability**

The data that support the findings of this study and all other relevant data are available from the corresponding author upon request.

**References**

1.  Mannhart, J., & Schlom, D. G. Oxide interfaces--an opportunity for electronics. *Science* **327**, 1607–1611 (2010).

2.  Hwang, H. Y., Iwasa, Y., Kawasaki, M., Keimer, B., Nagaosa, N., & Tokura, Y. Emergent phenomena at oxide interfaces. *Nat. Mater.* **11**, 103–113 (2012).

3.  Yu, P., Chu, Y. H., & Ramesh, R. Oxide interfaces: Pathways to novel phenomena. *Mater. Today* **15**, 320–327 (2012).

4.  Chakhalian, J., Millis, a. J., & Rondinelli, J. Whither the oxide interface. *Nat. Mater.* **11**, 92–94 (2012).

5.  Anderson, P. W., & Blount, E. I. Symmetry considerations on martensitic transformations: "Ferroelectric" metals? *Phys. Rev. Lett.* **14**, 532–532 (1965).

6.  Fei, Z. *et al.* Ferroelectric switching of a two-dimensional metal. *Nature* **560**, 336–339 (2018).

7.  Sharma, P. *et al.* A room-temperature ferroelectric semimetal. *Sci. Adv.* **5**, eaax5080 (2019).

8.  Zhou, W. X., & Ariando, A. Review on ferroelectric/polar metals. *Jpn. J. Appl. Phys.* **59**, SI0802 (2020).

9.  Kim, T. H. *et al.* Polar metals by geometric design. *Nature* **533**, 68–72 (2016).

10. Kolodiazhnyi, T., Tachibana, M., Kawaji, H., Hwang, J., & Takayama-Muromachi, E. Persistence of ferroelectricity in $BaTiO_3$ through the insulator-metal transition. *Phys. Rev. Lett.* **104**, 147602 (2010).

11. Dong, M. *et al.* Polar metallicity controlled by epitaxial strain engineering. *Adv. Sci.* 2408329 (2024).

12. Lei, S. *et al.* Observation of quasi-two-dimensional polar domains and ferroelastic switching in a metal, $Ca_3Ru_2O_7$. *Nano Lett.* **18**, 3088–3095 (2018).

13. Sakai, H. *et al.* Critical enhancement of thermopower in a chemically tuned polar semimetal $MoTe_2$. *Sci. Adv.* **2**, e1601378 (2016).

14. Ohtomo, A., & Hwang, H. Y. A high-mobility electron gas at the $LaAlO_3/SrTiO_3$ heterointerface. *Nature* **427**, 423–426 (2004).

15. Okamoto, S., & Millis, A. J. Electronic reconstruction at an interface between a Mott insulator and a band insulator. *Nature* **428**, 630–633 (2004).

16. Gozar, A. *et al.* High-temperature interface superconductivity between metallic and insulating copper oxides. *Nature* **455**, 782–785 (2008).

17. Wu, J. *et al.* Anomalous independence of interface superconductivity from carrier density. *Nat. Mater.* **12**, 877–881. (2013).

18. Shen, J. Y. *et al.* Reentrance of interface superconductivity in a high-$T_c$ cuprate heterostructure. *Nat. Commun.* **14**, 7290 (2023).

19. Reyren, N. *et al.* Superconducting interfaces between insulating oxides. *Science* **317**, 1196–1199 (2007).

20. Caviglia, A. D. *et al.* Electric field control of the $LaAlO_3/SrTiO_3$ interface ground state. *Nature* **456**, 624–627. (2008).

21. Wang, Q. Y. *et al.* Interface-induced high-temperature superconductivity in single unit-cell FeSe films on $SrTiO_3$. *Chin. Phys. Lett.* **29**, 037402 (2012).

22. Liu, C. *et al.* Two-dimensional superconductivity and anisotropic transport at $KTaO_3$(111) interfaces. *Science* **371**, 716–721 (2021).

23. Hua, X. *et al.* Superconducting stripes induced by ferromagnetic proximity in an oxide heterostructure. *Nat. Phys.* **20**, 957–963 (2024).

24. Chen, Z. *et al.* Electric field control of superconductivity at the $LaAlO_3/KTaO_3$(111) interface. *Science* **372**, 721–724 (2021).

25. Chen, Z. *et al.* Two-dimensional superconductivity at the $LaAlO_3/KTaO_3$(110) heterointerface. *Phys. Rev. Lett.* 126, 026802 (2021).

26. Arnault, E. G. *et al.* Anisotropic superconductivity at $KTaO_3$(111) interfaces. *Sci. Adv.* **9**, eadf1414 (2023).

27. Chen, X. *et al.* Orientation-dependent electronic structure in interfacial superconductors $LaAlO_3/KTaO_3$. *Nat. Commun.* **15**, 7704 (2024).

28. Ren, T. *et al.* Two-dimensional superconductivity at the surfaces of $KTaO_3$ gated with ionic liquid. *Sci. Adv.* **8**, eabn4273 (2022).

29. Mallik, S. *et al.* Superfluid stiffness of a $KTaO_3$-based two-dimensional electron gas. *Nat. Commun.* **13**, 4625 (2022).

30. Liu, C. *et al.* Tunable superconductivity and its origin at $KTaO_3$ interfaces. *Nat. Commun.* **14**, 951 (2023).

31. Zhang, G. *et al.* Spontaneous rotational symmetry breaking in $KTaO_3$ heterointerface superconductors. *Nat. Commun.* 14, 3046 (2023).

32. Müller, K. A. & Burkhard, H. $SrTiO_3$: an intrinsic quantum paraelectric below 4 K. *Phys. Rev. B* **19**, 3593–3602 (1979).

33. Fujii, Y. & Sakudo, T. Dielectric and optical properties of $KTaO_3$. *J. Phys. Soc. Jpn.* **41**, 888-893 (1976).

34. Rischau, C. W. *et al.* A ferroelectric quantum phase transition inside the superconducting dome of $Sr_{1-x}Ca_xTiO_{3-\delta}$. *Nat. Phys.* **13**, 643–648 (2017).

35. Jindal, A. *et al.* Coupled ferroelectricity and superconductivity in bilayer $T_d$-$MoTe_2$. *Nature* **613**, 48–52 (2023).

36. Denev, S. A., Lummen, T. T. A., Barnes, E., Kumar, A., & Gopalan, V. Probing ferroelectrics using optical second harmonic generation. *J. Am. Ceram. Soc.* **94**, 2699–2727 (2011).

37. Yang, Q. *et al.* Ferroelectricity in layered bismuth oxide down to 1 nanometer. *Science* **379**, 1218–1224 (2023).

38. Sun, H. *et al.* Nonvolatile ferroelectric domain wall memory integrated on silicon. *Nat. Commun.* **13**, 4332. (2022).

39. Io, W. F. *et al.* Direct observation of intrinsic room-temperature ferroelectricity in 2D layered $CuCrP_2S_6$. *Nat. Commun.* **14**, 7304. (2023).

40. Gruverman, A., Alexe, M., & Meier, D. Piezoresponse force microscopy and nanoferroic phenomena. *Nat. Commun.* **10**, 1661 (2019).

41. Lu, N. *et al.* Electric-field control of tri-state phase transformation with a selective dual-ion switch. *Nature* **546**, 124–128 (2017).

**Acknowledgements**

This work was supported by National Key R&D Program of China (2023YFA1406400 to J.W.), Research Center for Industries of the Future (RCIF project No. WU2023C001 to J.W.) at Westlake University, the National Natural Science Foundation of China (Grant No. 12174318 to J.W.), the Zhejiang Provincial Natural Science Foundation of China (Grant No. XHD23A2002 to J.W.). We acknowledge the assistance provided by Dr. Qike Jiang of the Instrumentation and Service Center for Physical Sciences and the Instrumentation and Service Centers for Molecular Science at Westlake University.

**Author contributions**

The film synthesis and lithography were done by M.Z., STEM, SHG, PFM and electrical transport measurements were done by M.D.D. and X.B.C., and the analysis and interpretation put forward by J.W.

**Competing interests**

The authors declare no competing interests.

**Additional information**

Supplementary information is available in the online version of the paper. Correspondence and requests for materials should be addressed to J.W.

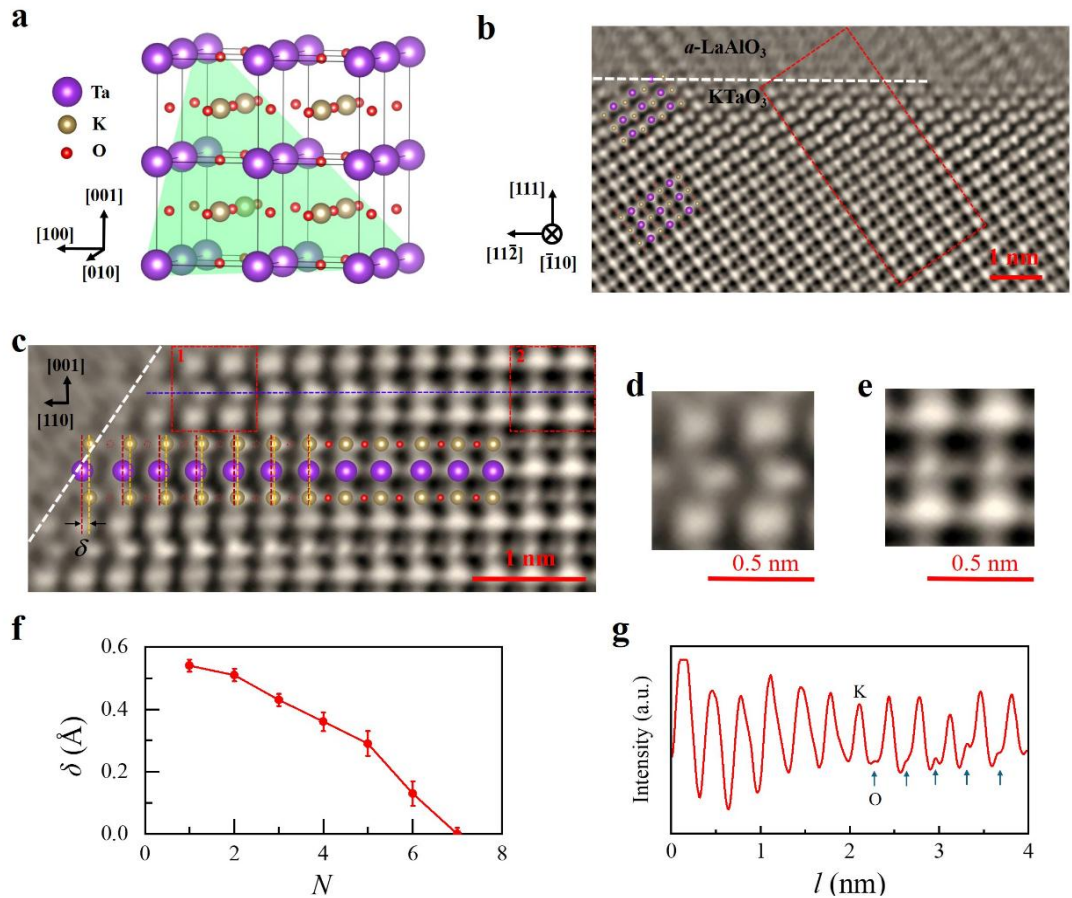

**Figure 1 | Atomic displacement and oxygen vacancies at LAO/KTO interface.** **a,** A schematic drawing of KTO crystalline structure. **b,** The iDPC-STEM image with atomic resolution. The dashed line denotes the boundary between LAO and KTO. **c,** The amplified view of atoms in the red box of panel **b**. The relative displacement between K and Ta atoms is denoted by $\delta$. $\delta$ is substantial (~ 0.54 Å) right at LAO/KTO interface and quickly approaches zero for atoms distant from the interface. Concomitantly, some oxygen atoms are clearly missing at the interface (open circles). **d**, and **e**, The amplified view of atoms in box 1 and 2 of panel **c**, respectively. The atomic structure close to and far away from the interface show remarkable differences. **f**, The measured $\delta$ as a function of $N$, where $N$ is the number of Ta-O plane counting from the interface. **g**, The line profile (corresponding to the purple dashed line in panel **c**) show the locations of oxygen atoms, which is missing in the vicinity of the interface.

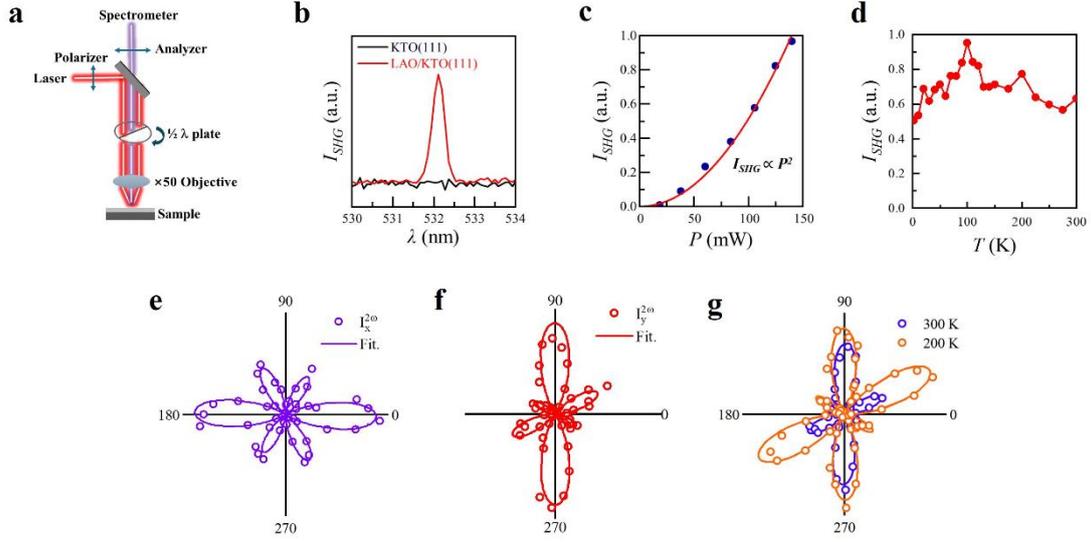

**Figure 2 │ Electric polarization of LAO/KTO interface characterized by SHG signal. a**, A schematic drawing of SHG microscope. The half-wavelength plate can be continuously rotated to measure both $I_x^{2\omega}(\phi)$ and $I_y^{2\omega}(\phi)$, where $\phi$ is the angle between the polarization of the incident light and the KTO crystallographic [1$\bar{1}$0] direction. **b,** The wavelength of incident laser is 1064 nm and pronounced SHG signal peaked at 532 nm wavelength is manifested in the reflected beam from LAO/KTO(111), in stark contrast to that of KTO(111) substrate itself. **c**, The intensity of SHG signal $I_{SHG}$ is proportional to $P^2$, where $P$ is the power of the incident laser. **d**, $I_{SHG}$ persists at low temperatures. **e**, and **f**, $I_x^{2\omega}(\phi)$ and $I_y^{2\omega}(\phi)$ measured at 300 K, respectively. The data (open circles) is well fitted (solid curves) by the expressions based on *m* polar group. **g**, $I_x^{2\omega}(\phi)$ measured at 300 K and 200 K show similar angular dependence and augmented amplitude at lower temperature, indicating that electric polarization persists and strengthens at low temperatures.

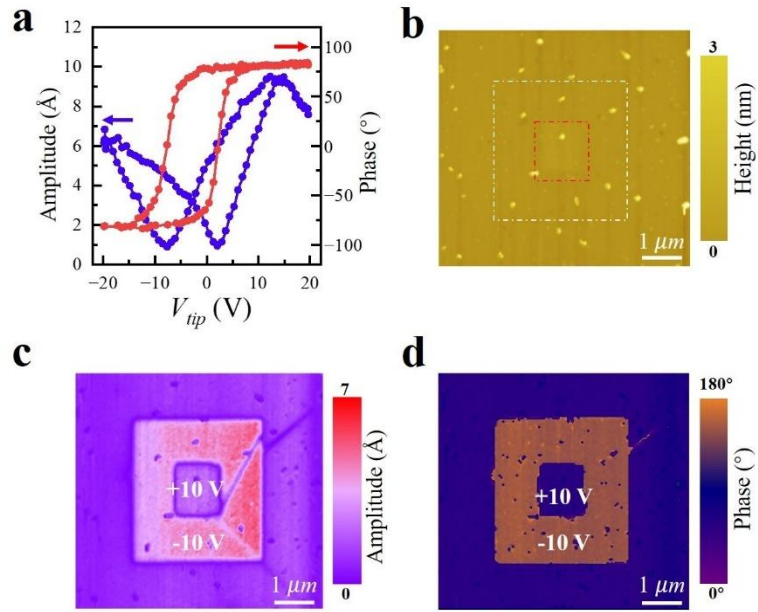

**Figure 3 | Ferroelectric domains created and manipulated by PFM.** **a,** The amplitude $A$ and phase $P$ of the piezoelectric response by ramping the voltage $V_{tip}$ applied between the PFM tip and the LAO/KTO sample. **b**, The atomic force microscopy shows flat surface of LAO/KTO. **c**, and **d**, The amplitude and phase of PFM manifest clear contrast for different areas taken after the writing of ferroelectric domains, evidencing the switchability of ferroelectricity. A ±10 V voltage is applied on the PFM tip during scanning to locally flip ferroelectric domains. The 4 × 4 µm² area within the yellow dashed lines was scanned with -10 V voltage on the PFM tip and then the inner 2 × 2 µm² area within the dashed red lines was scanned with +10 V voltage.

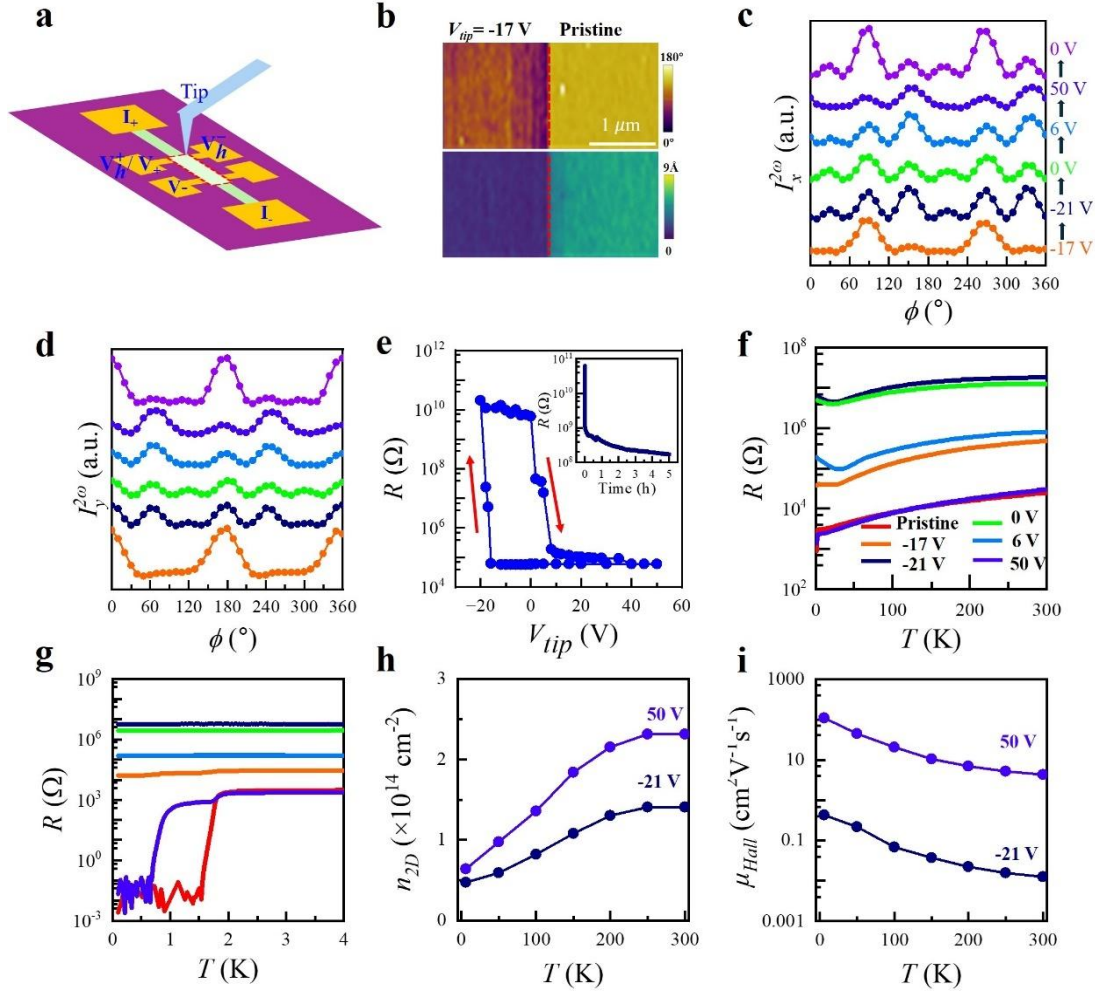

**Figure 4 | Strong coupling between ferroelectricity and superconductivity.** **a,** A schematic drawing of the experimental setup. The area inside the dashed lines was modulated by PFM tip scanning with variable tip voltage and then its longitudinal/Hall resistance was measured by a standard four-point measurement scheme. In this way, the correlation between ferroelectric hysteresis and electric transport can be elucidated. **b,** The boundary between the written (with $V_{tip}$ = -17 V) and pristine area is clearly visible on both the PFM amplitude and phase mapping, verifying that the ferroelectricity has been locally modulated. **c,** and **d,** The SHG signal $I_x^{2\omega}(\phi)$ and $I_y^{2\omega}(\phi)$ at 300 K corresponding to the cycling of $V_{tip}$. The sophisticated evolution of $I_x^{2\omega}(\phi)$ and $I_y^{2\omega}(\phi)$ with $V_{tip}$ reflects the motion and flipping of ferroelectric domains. **e,** The LAO/KTO longitudinal resistance manifests clear hysteresis behavior in response to PFM tip writings at 300 K. Remarkably, the change in resistance is gigantic and reaches more

than $10^5$ times for two saturated states of hysteresis loop. The inset shows the decay of $R$ as a function of time passed after the writing process. **f**, The corresponding $R(T)$ at different $V_{tip}$. $R(T)$ returns to that of the pristine state at the end of $V_{tip}$ loop after $V_{tip}$ = 50 V is applied, verifying that the change is non-volatile. **g**, $R(T)$ at low temperatures manifest the demise and reentrance of superconductivity during the cycling of $V_{tip}$, evidencing the strong coupling between superconductivity and ferroelectricity. **h**, The carrier density $n_{2D}$ determined by the Hall effect measurements changes by a factor of 1.4 for low and high resistance states ($V_{tip}$ = -21 V and 50 V, respectively). **i**, In contrast, the carrier mobility $\mu_{Hall}$ decreases 261 times from low to high resistance states.

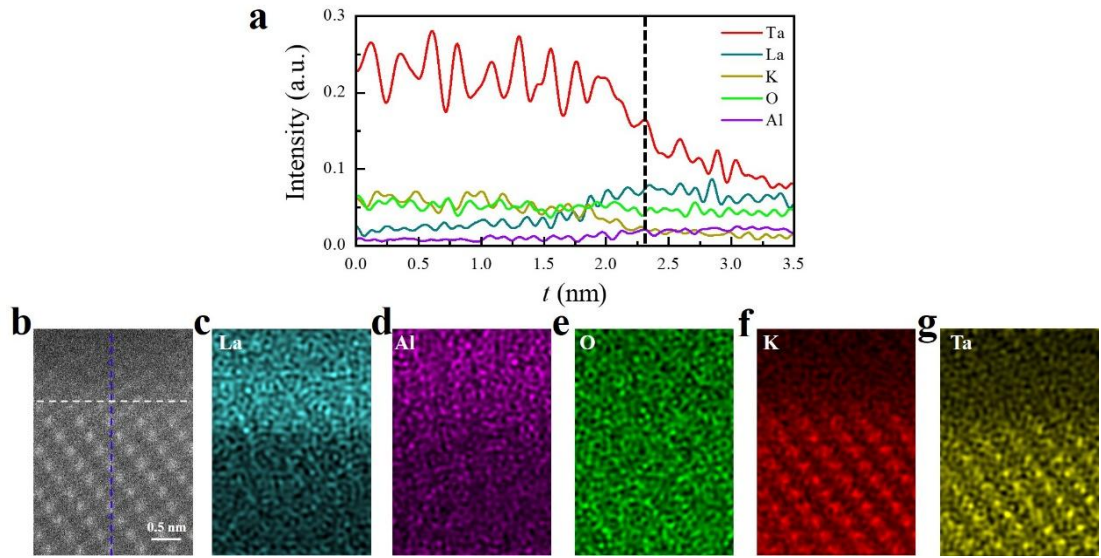

**Extended Data Figure 1 │ EDS-STEM mappings.** **a,** The line profile (corresponding to the purple dashed line in panel **b**) of different elements across the LAO/KTO interface. The location of the interface is indicated by the dashed line. **b**, The HAADF-STEM image. **c-g**, EDS mappings for the La, Al, O, Ta, and K elements, respectively. The interfacial cation interdiffusion is within 1 nm.

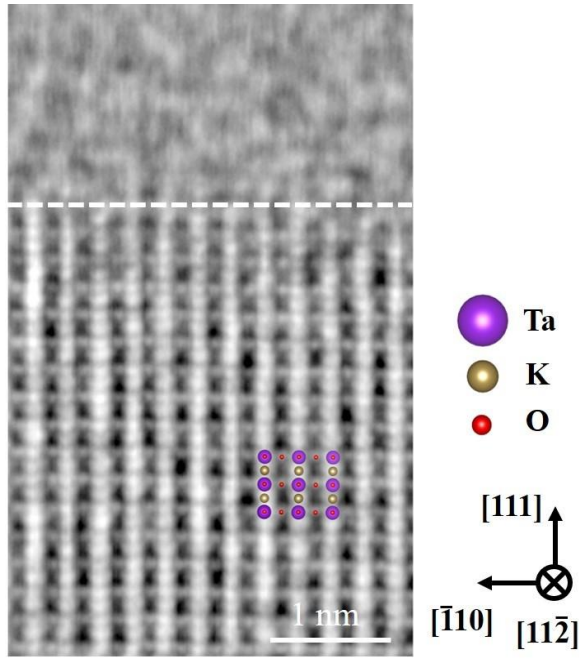

**Extended Data Figure 2 │ The iDPC-STEM image of LAO/KTO(111) with electrons incident along the KTO [11$\bar{2}$] direction.** The interface is denoted by the dashed line. Atomic displacement and oxygen vacancies, which are clearly visible for electrons incident along the KTO [$\bar{1}$10] direction (Fig. 1), are not distinguishable from this viewing angle. This evidences that the K-Ta displacement is entirely along KTO [110] direction and thus has no projection in the (11$\bar{2}$) plane. Concomitantly, oxygen vacancies residing at the K-O plane (Fig. 1), overlap with K atoms in position and hence is invisible from this viewing angle. Meanwhile, there is no sign of oxygen vacancies in the Ta-O plane at the interface.

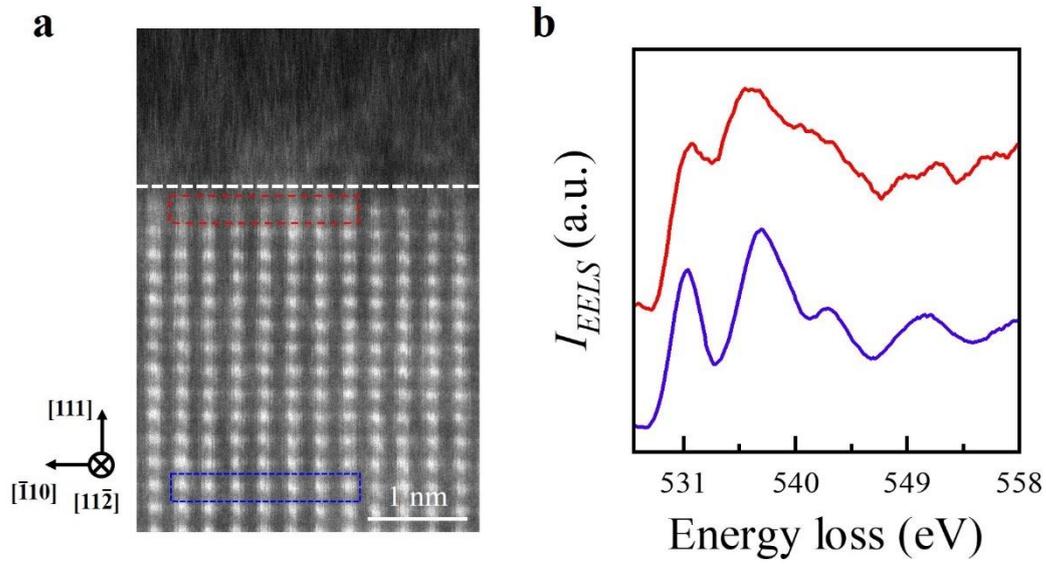

**Extended Data Figure 3 │ EELS spectra of oxygen K edge.** **a,** The HAADF-STEM image of LAO/KTO(111) sample. **b**, The EELS spectrums for area close to (red curve corresponding to the red box in panel **a**) and far away (blue curve, blue box in panel **a**) from the interface. The oxygen K edge absorption is much weaker for the red curve, implying a high density of oxygen vacancies at the interface.

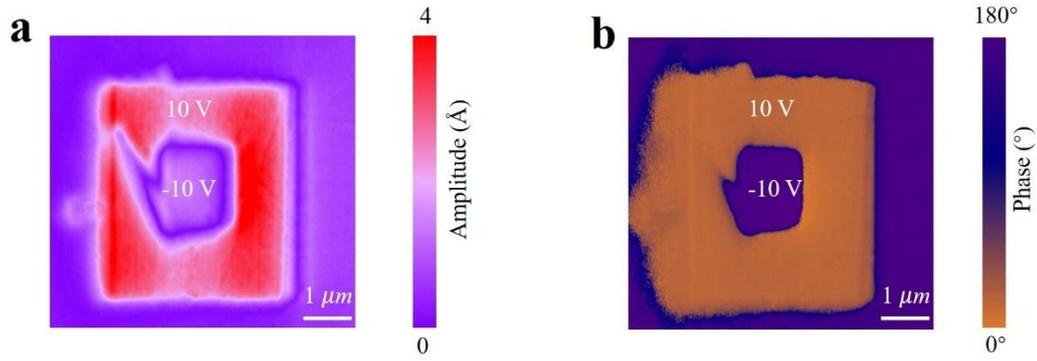

**Extended Data Figure 4 │ Another example of ferroelectric pattern written by PFM.** The inner and outer squares are 2 × 2 and 4 × 4 μm$^2$ written by applying +10 V and -10 V to the PFM tip respectively.

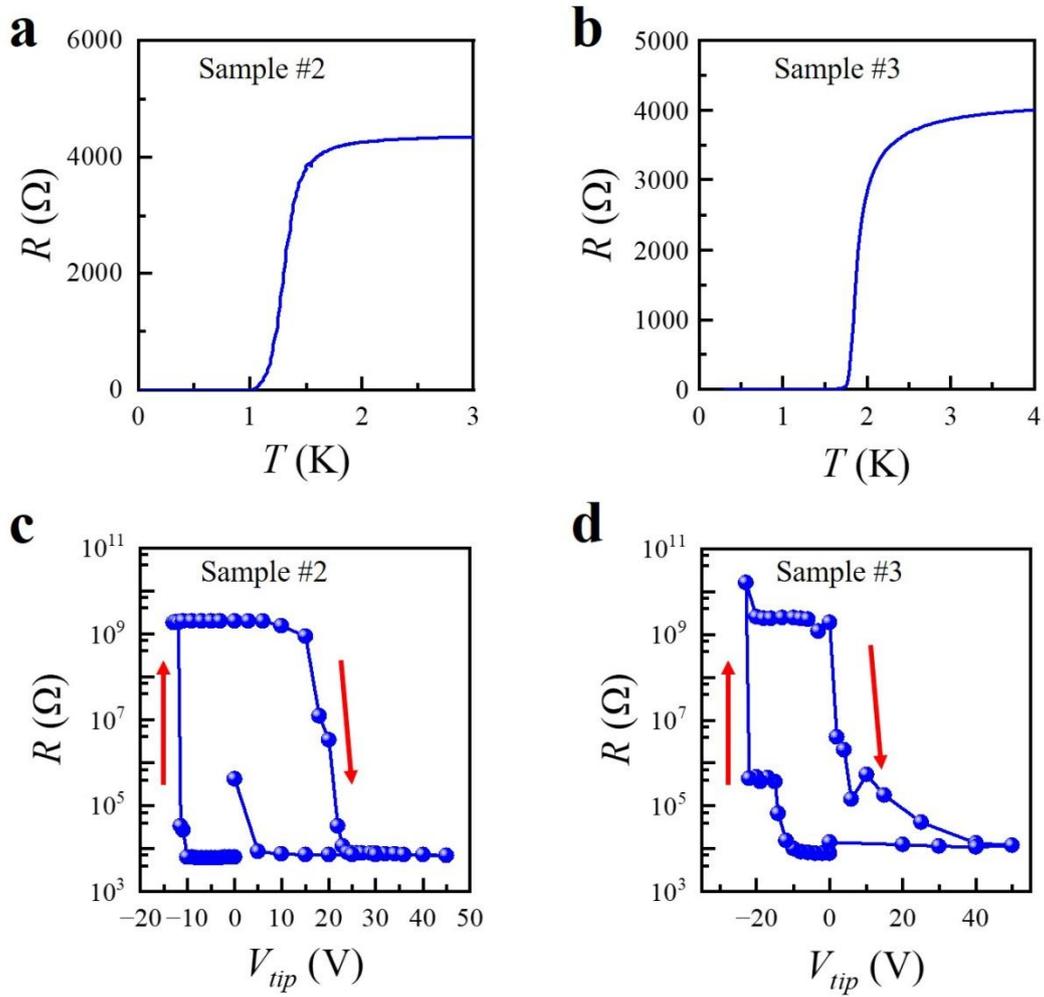

**Extended Data Figure 5 │ The superconductivity and hysteresis loops of another two LAO/KTO(111) samples.** **a,** and **b**, $R(T)$ manifest clear superconducting transitions but with different $T_c$ for sample #2 and #3. **c**, and **d**, Despite of differences in $T_c$, both samples show hysteresis behavior as the measured area are modulated by the PFM tip. These results are qualitatively the same to those in Fig. 4, showing that the strong coupling between ferroelectricity and superconductivity is universal for LAO/KTO(111) samples.